\documentclass[aps,prd,onecolumn,groupedaddress,showpacs,nofootinbib,amssymb]{revtex4}
\usepackage[dvips]{graphicx}
\usepackage{amssymb}
\usepackage{amsmath}
\usepackage{graphicx,,color}
\usepackage{amsfonts}
\usepackage{bm}
\usepackage{cancel}
\usepackage{comment}
\usepackage{floatflt}

\newcommand\be{\begin{equation}}
\newcommand\ee{\end{equation}}
\newcommand\nn{\nonumber \\}
\newcommand\e{\mathrm{e}}

\allowdisplaybreaks[4]

\begin{document}

\tolerance=5000

\title{Formalizing Anisotropic Inflation in Modified Gravity}
\author{S.~Nojiri,$^{1,2}$}
\email{nojiri@gravity.phys.nagoya-u.ac.jp}
\author{S.D.~Odintsov,$^{3,4}$}
\email{odintsov@ice.cat}
\author{V.K.~Oikonomou,$^{5}$}
\email{v.k.oikonomou1979@gmail.com;voikonomou@auth.gr}
\author{A.~Constantini$^{4}$}
\affiliation{$^{1)}$ Department of Physics, Nagoya University, Nagoya 464-8602, Japan \\
$^{2)}$ Kobayashi-Maskawa Institute for the Origin of Particles
and the Universe, Nagoya University, Nagoya 464-8602, Japan \\
$^{3)}$ ICREA, Passeig Luis Companys, 23, 08010 Barcelona, Spain\\
$^{4)}$ Institute of Space Sciences (ICE, CSIC) C. Can Magrans
s/n, 08193 Barcelona, Spain\\
$^{5)}$ Department of Physics, Aristotle University of
Thessaloniki, Thessaloniki 54124, Greece}

\begin{abstract}
Motivated by the fact that the pre-inflationary era may evolve in
an exotic way, in this work we formalize anisotropic evolution in
the context of modified gravity, focusing on pre-inflationary and
near the vicinity of the inflationary epochs. We specialize on
specific metrics like Bianchi and Taub and we formalize the
inflationary theory in vacuum $F(R)$ gravity, in $F(R)$ gravity
with an extra scalar field and in Gauss-Bonnet gravity. We discuss
the qualitative effects of the anisotropies on the evolution of
the Universe and also we consider several specific solutions, like
the de Sitter solution, in both the isotropic and anisotropic
contexts. Furthermore, several exotic modified gravity
cosmological solutions, like the ones which contain finite time
singularities, are also discussed in brief.
 \end{abstract}


\maketitle

\def\be{\begin{equation}}
\def\ee{\end{equation}}
\def\nn{\nonumber \\}
\def\e{\mathrm{e}}

\section{Introduction}

The Standard Cosmological Model provides a consistent general
picture of the Universe, at least after the inflationary era, and
it is entirely based on the standard Einstein-Hilbert gravity, on
the Standard model of particles and on the adoption of a
Friedmann-Robertson-Walker spacetime metric to describe the
geometry of the Universe. However, it is a well-known fact that,
at very primordial instants, at a pre-inflationary stage, the
physical description of the Universe can be more complicated than
a simple isotropic model and additional effects from modified and
quantum gravity can also become important. What is a well accepted
fact in cosmology is that the Universe when the presumed
inflationary era commenced, was four dimensional and it was
basically described by classical physics. The inflationary era
itself \cite{inflation1,inflation2,inflation3,inflation4} can
either be described by a single scalar field, in a purely general
relativistic (GR) framework. However, in the case that inflation
is controlled by the inflaton, many shortcomings of the theory
exist, such as the way that the inflaton couples to the Standard
Model particles is arbitrary, and so on. In conjunction with the
shortcomings of the scalar field description of the inflationary
era, the dark energy era described by a scalar field can also be
problematic, mainly due to the fact that observations
\cite{Planck:2018vyg} allow the dark energy equation of state
parameters to take values that may describe slightly a phantom
evolution and specifically $\omega_{DE}=-0.957 \pm 0.080$ and
hence the minimum observationally allowed value of the dark energy
equation of state parameter is $\omega_{DE}=-1.037$
\cite{Planck:2018vyg}. Although the Planck data allow an exact de
Sitter (cosmological constant) and quintessential dark energy era,
it also allows a phantom dark energy era. This perspective is
problematic in the context of simple GR, since a phantom evolution
can only be realized using a phantom scalar. The latter are in
general non appealing descriptions. Thus GR descriptions with
scalar fields have several conceptual issues that cannot be
appropriately resolved easily.

Modified gravity
\cite{reviews1,reviews2,reviews3,reviews4,reviews5} on the other
hand provide a consistent theoretical framework in the context of
which, both inflation and late-time acceleration can be described
in a unified way, see the pioneer work \cite{Nojiri:2003ft} for
this topic and also Refs.
\cite{Appleby:2009uf,Artymowski:2014gea,Yashiki:2020naf,Ellis:2020krl,Chen:2022zkc,Nojiri:2007as,Nojiri:2007cq,Cognola:2007zu,Nojiri:2006gh,Appleby:2007vb,Elizalde:2010ts,Odintsov:2020nwm,Sa:2020fvn}
for later developments on this topic. In fact, in the context of
modified gravity, the dark energy era can turn into phantom, and
at the same time be compatible with the Planck data, without
relying to phantom scalars, see Ref. \cite{Oikonomou:2022wuk} for
a several examples of this sort.

Now about the pre-inflationary era, this is a mystery for
scientists, and for the moment it is experimentally inaccessible
to us in a direct way. It is however possible that the
pre-inflationary epoch, described by some version of M-theory, may
leave its imprints on the low-energy effective inflationary
theory, and this can be done in several ways, for example some
higher order curvature terms, Einstein-Gauss-Bonnet terms or even
via some exotic evolution scenarios, like pre-inflationary bounces
\cite{Odintsov:2021urx}, other effects \cite{Bamonti:2021jmg} or
even via effects of anisotropic evolution. Thus we can have both
UV-corrections on the inflationary era or even anisotropy could be
generated during the pre-inflation era, leading to anisotropic
inflation. In this paper we shall consider the effects of the
second perspective.

In this paper we shall adopt the last line of research and we
shall formalize anisotropic inflation
\cite{Chen:2021nkf,Do:2021lyf,Sadeghi:2021egp,Hiramatsu:2020jes,Do:2020ler,Gong:2019hwj,Fujita:2018zbr,Ito:2017bnn,Lahiri:2016jqv,Ito:2015sxj,Blanco-Pillado:2015dfa,Naruko:2014bxa,
Emami:2013bk,Watanabe:2010bu,Kanno:2010nr,Watanabe:2010fh,Dulaney:2010sq,Barrow:2009gx,Rothman:1986gg}
in the context of modified gravity. The most natural
generalization of the isotropic FRW geometry is provided by the
Bianchi metrics classification. In particular, the Bianchi type
VIII and IX cosmological model yield chaotic dynamics in ordinary
GR contexts, which constitutes a prototype for the asymptotic
behavior to the singularity of the generic inhomogeneous Universe
\cite{Belinsky:1982pk}. We aim to describe the features of the
Universe adopting the Bianchi IX cosmology, which is the most
general geometry allowed by the homogeneity constraint. The
relevance of the dynamics of Bianchi models consists in the role
these geometries could have played in a very primordial Universe,
before the inflationary phase. For a quantum discussion on the
Bianchi IX model see \cite{Wilson-Ewing:2017vju,
Antonini:2018gdd}. The model has been extensively studied in GR by
C.Misner \cite{Misner:1969hg,Misner:1969ae}  but few analysis were
made in modified Gauss-Bonnet and $F(R)$ gravity.

The paper is thematically organized in three distinct sections,
each of which is devoted to studying anisotropic inflation in
distinct modified gravities. Specifically, in section II we
present the formalism of anisotropic inflation in the context of
vacuum $F(R)$ gravity. In section III we present the $F(R)$
gravity with an extra scalar field anisotropic inflation. In
section IV we consider the anisotropic Gauss-Bonnet inflation and
in section we formalize anisotropic Gauss-Bonnet inflation in Taub
spacetimes. Finally the conclusions follow in the end of the
article.

\section{Anisotropic Inflation in the Case of $F(R)$ Gravity}

In this section we shall formalize the anisotropic inflation in
the context of vacuum Jordan frame $F(R)$ gravity. Firstly let us
fix the metric, so we assume the following homogeneous and
anisotropic metric,
\begin{align}
\label{an1} ds^2 = - dt^2 + a(t)^2 \sum_{i=1}^3 \e^{2\beta_i(t)}
\left( dx^i \right)^2 \,  .
\end{align}
Let the average of $\beta_i(t)$ be $\bar\beta(t)$, $\bar\beta(t)
\equiv \frac{1}{3}\sum_{i=1}^3 \beta_i(t)$. By redfining $a(t)$
and $\beta_i(t)$ as $a(t)\to a(t) + \bar\beta(t)$ and
$\beta_i(t)\to \beta_i(t) - \bar\beta(t)$, we obtain
\begin{align}
\label{an3}
0 = \sum_{i=1}^3 \beta^i \, .
\end{align}
and therefore
\begin{align}
\label{an2}
0 = \sum_{i=1}^3 {\dot\beta}^i \, .
\end{align}
In the following, we assume Eqs.~(\ref{an3}) and (\ref{an2}).

Then, the connections are given by,
\begin{align}
\label{conn}
\Gamma^t_{ij} =  a^2 \e^{2\beta_i} \left( H + {\dot\beta}^i \right) \delta_{ij} \, , \quad
\Gamma^i_{tj}=\Gamma^i_{jt}= \left( H + {\dot\beta}^i \right) \delta^i_{\ j} \, , \quad
\mbox{other components}=0 \, .
\end{align}
Since the energy-momentum tensor is given by,
\begin{align}
\label{rhop}
T_{00}=-g_{00}\rho\ ,\quad T_{ij}=pg_{ij}\, ,
\end{align}
the conservation law is given by,
\begin{align}
\label{conslaw}
0 =&\, \nabla^\mu T_{\mu t} = - \dot\rho - \sum_{i,j=1}^3 a^{-2} \e^{-2\beta_i} \delta^{ij}  a^2 \e^{2\beta_i} \left( H + {\dot\beta}^i \right) \delta_{ij} \rho
 - a^{-2} \e^{-2\beta_i} \delta^{ij} \left( H + {\dot\beta}^i \right) \delta^k_{\ j} a^2 \e^{2\beta_i} \delta_{kj}  p \nonumber \\
=&\, - \left\{ \dot\rho + 3 H \left( \rho + p \right) \right\}\, .
\end{align}
Because we define,
\begin{align}
\label{Ricci}
R_{\mu\nu}=-\Gamma^\rho_{\mu\rho,\nu}
+ \Gamma^\rho_{\mu\nu,\rho}
 - \Gamma^\eta_{\mu\rho}\Gamma^\rho_{\nu\eta}
+ \Gamma^\eta_{\mu\nu}\Gamma^\rho_{\rho\eta} \, ,
\end{align}
we find that the components of the Ricci tensor are,
\begin{align}
\label{Ricci2}
R_{tt} =&\, - 3 \dot H - \sum_{i=1}^3 \left( H + {\dot\beta}^i \right)^2 = -3 \dot H - 3 H^2  - \sum_{i=1}^3 \left({\dot\beta}^i\right)^2 \, , \nonumber \\
R_{ij} =&\, a^2 \e^{2\beta_i} \left( \dot H + {\ddot\beta}^i \right) \delta_{ij} + 2 a^2 \e^{2\beta_i} \left( H + {\dot\beta}^i \right)^2 \delta_{ij}
 - 2 a^2 \e^{2\beta_i} \left( H + {\dot\beta}^i \right)^2 \delta_{ij} + a^2 \e^{2\beta_i} \left( H + {\dot\beta}^i \right) \delta_{ij} 3 H \nonumber \\
=&\, a^2 \e^{2\beta_i} \left( \dot H + 3 H^2 + {\ddot\beta}^i + 3 H {\dot\beta}^i\right) \delta_{ij} \nonumber \\
R=&\, 6 \dot H + 12 H^2 + \sum_{i=1}^3
\left({\dot\beta}^i\right)^2 \, ,
\end{align}
where we have used $0 = \sum_{i=1}^3 {\ddot\beta}^i = \frac{d}{dt}
\sum_{i=1}^3 {\dot\beta}^i $. The field equations for $F(R)$
gravity are given by,
\begin{equation}
\label{JGRG13} G^F_{\mu\nu} \equiv \frac{1}{2}g_{\mu\nu} F -
R_{\mu\nu} F_R - g_{\mu\nu} \Box F_R + \nabla_\mu \nabla_\nu F_R =
- \kappa^2 T_{\mu\nu}\, ,
\end{equation}
where $F_R \equiv \frac{dF(R)}{dR}$. Since $\Box F_R \equiv
\frac{1}{\sqrt{-g}} \partial_\mu \left( g^{\mu\nu} \sqrt{-g}
\partial_\nu F_R \right)= - \left( \frac{d^2 F_R}{dt^2} + 3 H
\frac{d F_R}{dt} \right)$, we find that the $tt$ component of
(\ref{JGRG13}) is given by,
\begin{align}
\label{FRB1}
&\, - \frac{1}{2} F +\left( 3 \dot H + 3 H^2  + \sum_{i=1}^3 \left({\dot\beta}^i\right)^2 \right) F_R - \left( \frac{d^2 F_R}{dt^2} + 3 H \frac{d F_R}{dt} \right)
+ \frac{d^2 F_R}{dt^2} \nonumber \\
=&\, - \frac{1}{2} F +\left( 3 \dot H + 3 H^2  + \sum_{i=1}^3 \left({\dot\beta}^i\right)^2 \right) F_R - 3 H \frac{d F_R}{dt} = - \kappa^2 \rho \, ,
\end{align}
and $ij$ component yields the following,
\begin{align}
\label{FRB2}
&\, \frac{1}{2} F  -\left(  \dot H + 3 H^2 + {\ddot\beta}^i + 3 H {\dot\beta}^i\right) F_R + \left( \frac{d^2 F_R}{dt^2} + 3 H \frac{d F_R}{dt} \right)
 - \left( H + {\dot\beta}^i \right)  \frac{d F_R}{dt} \nonumber \\
= &\, \frac{1}{2} F  -\left(  \dot H + 3 H^2 + {\ddot\beta}^i + 3 H {\dot\beta}^i\right) F_R + \frac{d^2 F_R}{dt^2} + \left( 2H - {\dot\beta}^i \right) \frac{d F_R}{dt} = - \kappa^2 p \, ,
\end{align}
which gives,
\begin{align}
\label{FRB3}
&\, \frac{1}{2} F  -\left(  \dot H + 3 H^2 \right) F_R + \frac{d^2 F_R}{dt^2} + 2H \frac{d F_R}{dt} = - \kappa^2 p \, , \\
\label{FRB4} &\, -\left( {\ddot\beta}^i + 3 H {\dot\beta}^i\right)
F_R - {\dot\beta}^i \frac{d F_R}{dt} = 0 \, .
\end{align}
The above equation, namely Eq.~(\ref{FRB3}), can be obtained by
summing up Eq.~(\ref{FRB2}) with respect to $i$ and using
(\ref{an2}) and also the fact that $0 = \sum_{i=1}^3
{\ddot\beta}^i$. We obtain Eq.~(\ref{FRB4}) by subtracting
Eq.~(\ref{FRB2}) and Eq.~(\ref{FRB3}).

We now check the consistency between the conservation law
(\ref{conslaw}) with the equations (\ref{FRB1}), (\ref{FRB3}), and
(\ref{FRB4}), and we get,
\begin{align}
\label{conslaw1}
 - \kappa^2 &\, \left\{ \dot\rho + 3 H \left( \rho + p \right) \right\} \nonumber \\
=&\, - \frac{1}{2} F_R \left( 6 \ddot H + 24 H \dot H + 2 \sum_{i=1}^3 {\dot\beta}^i {\ddot\beta}^i \right)
+\left( 3 \ddot H + 6 H \dot H  + 2 \sum_{i=1}^3 {\dot\beta}^i {\ddot\beta}^i \right) F_R
+\left( 3 \dot H + 3 H^2  + \sum_{i=1}^3 \left({\dot\beta}^i\right)^2 \right) \frac{dF_R}{dt} \nonumber \\
&\, - 3 \dot H \frac{d F_R}{dt} - 3 H \frac{d^2 F_R}{dt^2}
+ 3 H \left\{  - \frac{1}{2} F +\left( 3 \dot H + 3 H^2  + \sum_{i=1}^3 \left({\dot\beta}^i\right)^2 \right) F_R - 3 H \frac{d F_R}{dt} \right. \nonumber \\
&\, \left. + \frac{1}{2} F  -\left(  \dot H + 3 H^2 \right) F_R + \frac{d^2 F_R}{dt^2} + 2H \frac{d F_R}{dt} \right\} \nonumber \\
=&\, - \frac{1}{2} F_R \left(  24 H \dot H + 2 \sum_{i=1}^3 {\dot\beta}^i {\ddot\beta}^i \right)
+\left( 6 H \dot H  + 2 \sum_{i=1}^3 {\dot\beta}^i {\ddot\beta}^i \right) F_R
+\left( 3 H^2  + \sum_{i=1}^3 \left({\dot\beta}^i\right)^2 \right) \frac{dF_R}{dt} \nonumber \\
&\, + 3 H \left\{ \left( 3 \dot H + 3 H^2  + \sum_{i=1}^3 \left({\dot\beta}^i\right)^2 \right) F_R - 3 H \frac{d F_R}{dt}
 -\left(  \dot H + 3 H^2 \right) F_R + 2H \frac{d F_R}{dt} \right\} \nonumber \\
=&\, \left\{ \left( - 12 + 6 + 9 - 3\right) H\dot H + \left( 9 - 9 \right) H^2 \right\} F_R
+ \left( 3 -9 + 6\right) H^2 \frac{dF_R}{dt} \nonumber \\
&\, + \sum_{i=1}^3 {\dot\beta}^i \left\{ \left( - 1 + 2 \right) {\ddot\beta}^i F_R
+ 3 H {\dot\beta}^i F_R + {\dot\beta}^i \frac{dF_R}{dt} \right\} \nonumber \\
=&\, 0 \, .
\end{align}
Therefore there is no contradiction in the equations which we
obtained, these are consistent.

Eq.~(\ref{FRB4}) can be integrated with respect to ${\dot
\beta}^i$ to yield,
\begin{align}
\label{FRB5} {\dot \beta}^i = \frac{C^i a^{-3}}{F_R}\, ,
\end{align}
where the free parameters $C^i$'s are constants. By integrating
${\dot \beta}^i$ with respect $t$, we obtain $\beta^i$ and also
Eq.~(\ref{an2}) indicates that,
\begin{align}
\label{FRB6}
0 = \sum_{i=1}^3 C^i \, .
\end{align}
In order for the late Universe to satisfy $a\to \infty$ and $R\to
0$, and in order for it to become isotropic, we must require
$\frac{a^{-3}}{F_R} \to 0$.

By combining Eqs. (\ref{FRB1}) and (\ref{FRB3}) and eliminating
${\dot \beta}^i$ by using (\ref{FRB5}), we obtain,
\begin{align}
\label{FRB7}
\left( 2 \dot H + \frac{C^2 a^{-6}}{{F_R}^2} \right) F_R - H \frac{d F_R}{dt}
+ \frac{d^2 F_R}{dt^2} = - \kappa^2 \left( \rho + p \right)\, ,
\end{align}
with
\begin{align}
\label{FRB8}
C^2 \equiv \sum_{i=1} {C^i}^2 \, .
\end{align}
We may assume that the matter is composed by perfect fluids with
constant EoS parameters $w_i$,
\begin{align}
\label{FRB8} \rho + p = \sum_i \left( 1 + w_i \right) \rho_0^i
a^{-3\left( 1 + w_i \right)} \, ,
\end{align}
where the parameters $\rho_0^i$'s are constants. If we give the
time evolution of the scale factor $a$ as $a=a(t)$, and therefore
$H=H(t)$ and $\dot H = \dot H(t)$, Eq.~(\ref{FRB7}) becomes a
non-linear differential equation for $F_R$. Then by solving
Eq.~(\ref{FRB7}), we may find the time dependence of $F_R$ as
$F_R=F_R(t)$. On the other hand, since the scalar curvature $R$ is
given by,
\begin{align}
\label{FRB9} R=R(t)=12 H(t)^2 + 6 \dot H(t) + \frac{C^2
a^{-6}}{{F_R}^2} \, ,
\end{align}
if we delete $t$ by combining $F_R=F_R(t)$ and $R=R(t)$, $F_R$ is
given by a function of $R$ and by integrating $F_R(R)$ with
respect to $R$, we find the form of $F(R)$.

We now consider the case without matter, so $\rho+p=0$. Then
Eq.~(\ref{FRB7}) takes the following form,
\begin{align}
\label{FRB10}
2 \dot H F_R - H \frac{d F_R}{dt}
+ \frac{d^2 F_R}{dt^2} = - \frac{C^2 a^{-6}}{F_R}\, .
\end{align}
During the inflationary era, that is, the era after the first
horizon crossing and much earlier than the reheating, the Hubble
rate $H$ is almost constant, $H=H_0$, which indicates that we may
neglect the first term in the left hand side of Eq.~(\ref{FRB10})
and $a=\e^{H_0 t}$. Then, we may write Eq.~(\ref{FRB10}) in the
following form,
\begin{align}
\label{FRB11}
 - H_0 \frac{d F_R}{dt} + \frac{d^2 F_R}{dt^2} = \e^{H_0t} \frac{d}{dt} \left( \e^{-H_0t} \frac{d F_R}{dt} \right) = - \frac{C^2 \e^{-6H_0t}}{F_R}\, .
\end{align}
In the late epoch of the inflationary era, but much before than
the reheating, the right hand side of (\ref{FRB11}) should be
small and can be treated as a perturbation. The leading behavior
can be obtained when the right hand side of Eq. (\ref{FRB11})
vanishes, and we find that,
\begin{align}
\label{FRB12} F_R = F_1 + F_2 \e^{H_0 t}\, ,
\end{align}
where $F_1$ and $F_2$ are constants, which may be determined by
the initial condition. When $F_2=0$ by an initial condition, the
solution of Eq. (\ref{FRB11}) including the leading correction is
given by,
\begin{align}
\label{FRB13}
F_R = F_1 - \frac{C^2 \e^{-6H_0t}}{42 {H_0}^2 F_1} \, .
\end{align}
On the other hand, when $F_2\neq 0$, the second term in (\ref{FRB12}) becomes dominant in the late-era and, we find,
\begin{align}
\label{FRB14}
F_R = F_1 + F_2 \e^{H_0 t} - \frac{C^2 \e^{-7H_0t}}{56 {H_0}^2 F_2}\, .
\end{align}
In the case that $F_2=0$ or $F_2\neq 0$, we find that
$\frac{a^{-3}}{F_R}$ decreases very rapidly and the Universe
becomes isotropic. Eq.~(\ref{FRB9}) also indicates that,
\begin{align}
\label{FRB15}
R \sim \left\{ \begin{array}{cc}
12{H_0}^2 + \frac{C^2 \e^{-6H_0 t}}{{F_1}^2} & \mathrm{when}\ F_2=0 \\
12{H_0}^2 + \frac{C^2 \e^{-8H_0 t}}{{F_2}^2} & \mathrm{when}\
F_2\neq 0
\end{array}
\right. \, ,
\end{align}
and consequently we find that,
\begin{align}
\label{FRB16}
F_R \sim \left\{ \begin{array}{cc}
F_1 - \frac{F_1}{42 {H_0}^2} \left( R - 12 {H_0}^2 \right)  & \mbox{when}\ F_2=0 \\
F_1 + F_2 \left\{ \frac{{F_2}^2\left( R - 12 {H_0}^2 \right)}{C^2} \right\}^{- \frac{1}{8}}
 - \frac{C^2}{56 {H_0}^2 F_2} \left\{ \frac{{F_2}^2\left( R - 12 {H_0}^2 \right)}{C^2} \right\}^\frac{7}{8} & \mbox{when}\ F_2\neq 0
\end{array}
\right. \, ,
\end{align}
By integrating $F_R$ with respect to $R$, we find the form of $F$
as follows,
\begin{align}
\label{FRB17}
F(R) \sim&\, \left\{ \begin{array}{cc}
F_0 + F_1 R - \frac{F_1}{42 {H_0}^2} \left( \frac{R^2}{2} - 12 {H_0}^2 R \right)  & \mbox{when}\ F_2=0 \\
F_0 + F_1 R + \frac{8C^2}{7F_2} \left\{ \frac{{F_2}^2\left( R - 12 {H_0}^2 \right)}{C^2} \right\}^\frac{7}{8}
 - \frac{8C^4}{840 {H_0}^2 {F_2}^3} \left\{ \frac{{F_2}^2\left( R - 12 {H_0}^2 \right)}{C^2} \right\}^\frac{15}{8} & \mbox{when}\ F_2\neq 0
\end{array}
\right. \nonumber \\
\sim&\, \left\{ \begin{array}{cc}
F_0 + F_1 R - \frac{F_1}{42 {H_0}^2} \left( \frac{R^2}{2} - 12 {H_0}^2 R \right)  & \mbox{when}\ F_2=0 \\
F_0 + F_1 R + \frac{8C^2}{7F_2} \left\{ \frac{{F_2}^2\left( R - 12 {H_0}^2 \right)}{C^2} \right\}^\frac{7}{8} & \mbox{when}\ F_2\neq 0
\end{array}
\right. \, ,
\end{align}
with a constant $F_0$ parameter. On the other hand, by using Eqs.
(\ref{FRB1}) or (\ref{FRB3}) with $\rho=p=0$, we can determine
$F(R)$. In order to check the consistency and to determine the
constant $F_0$, we consider  (\ref{FRB1}),
\begin{align}
\label{FRB18}
F = 2 \left( 3 \dot H + 3 H^2  + \sum_{i=1}^3 \left({\dot\beta}^i\right)^2 \right) F_R - 6 H \frac{d F_R}{dt} \, ,
\end{align}
and using Eqs. (\ref{FRB13}) and (\ref{FRB14}), we find,
\begin{align}
\label{FRB19}
F \sim&\, \left\{ \begin{array}{cc}
\left( 6 {H_0}^2 + \frac{2C^2 \e^{-6 H_0 t}}{{F_1}^2} \right) \left( F_1 - \frac{C^2 \e^{-6H_0t}}{42 {H_0}^2 F_1} \right)
 - \frac{6C^2 \e^{-6H_0t}}{7F_1} & \mbox{when}\ F_2=0 \\
\left( 6 {H_0}^2 + \frac{2C^2 \e^{-8 H_0 t}}{{F_2}^2} \right) \left(  F_1 + F_2 \e^{H_0 t} - \frac{C^2 \e^{-7H_0t}}{56 {H_0}^2 F_2} \right)
 -6 {H_0}^2 \left( F_2 H_0 \e^{H_0 t} + \frac{C^2 \e^{-7H_0t}}{8 {H_0}^2 F_2}\right) & \mbox{when}\ F_2\neq 0
\end{array}
\right. \nonumber \\
\sim&\, \left\{ \begin{array}{cc}
6 {H_0}^2 F_1 + \left( - \frac{1}{7} + 2 - \frac{6}{7} \right) \frac{C^2 \e^{-6H_0t}}{F_1} & \mbox{when}\ F_2=0 \\
6 {H_0}^2 F_1 + \frac{2C^2 F_1 \e^{-8 H_0 t}}{{F_2}^2} + \left( - \frac{3}{28} + 2 - \frac{3}{4} \right) \frac{C^2 \e^{-7H_0t}}{F_2} & \mbox{when}\ F_2\neq 0
\end{array}
\right. \nonumber \\
\sim&\, \left\{ \begin{array}{cc}
6 {H_0}^2 F_1 + \frac{C^2 \e^{-6H_0t}}{F_1} & \mbox{when}\ F_2=0 \\
6 {H_0}^2 F_1 + \frac{2C^2 F_1 \e^{-8 H_0 t}}{{F_2}^2} + \frac{8 C^2 \e^{-7H_0t}}{7 F_2} & \mbox{when}\ F_2\neq 0
\end{array}
\right. \nonumber \\
\sim&\, \left\{ \begin{array}{cc}
6 {H_0}^2 F_1 + \frac{C^2 \e^{-6H_0t}}{F_1} & \mbox{when}\ F_2=0 \\
6 {H_0}^2 F_1 + \frac{8 C^2 \e^{-7H_0t}}{7 F_2} & \mbox{when}\ F_2\neq 0
\end{array}
\right. \, ,
\end{align}
By using Eq. (\ref{FRB15}), we find that Eq.~(\ref{FRB19}) is
consistent with Eq. (\ref{FRB17}) if and only if the following
condition holds true,
\begin{align}
\label{FRB20}
F_0 =
\left\{ \begin{array}{cc}
 - \frac{54}{7} {H_0}^2 F_1  & \mbox{when}\ F_2=0 \\
 - 6 {H_0}^2 F_1 & \mbox{when}\ F_2\neq 0
\end{array}\right. \, ,
\end{align}

It is well-known that the $F(R)$ gravity can be rewritten as the scalar-tensor form by the scale transformation of the metric,
\begin{align}
\label{JGRG22}
g_{\mu\nu}\to \e^\frac{\varphi}{\sqrt{3}} g_{\mu\nu}\, ,\quad \varphi = -\frac{1}{\sqrt{3}}\ln F_R(A)\, ,
\end{align}
which gives the so-called Einstein frame action as follows,
\begin{align}
\label{JGRG23}
S_E =&\, \frac{1}{2\kappa^2}\int d^4 x \sqrt{-g} \left( R - \frac{1}{2}g^{\rho\sigma} \partial_\rho \varphi \partial_\sigma \varphi - V(\varphi)\right) \, ,\nonumber \\
V(\varphi) =&\, \e^\frac{\varphi}{\sqrt{3}} g\left(\e^{-\frac{\varphi}{\sqrt{3}}}\right) - \e^{\frac{2\varphi}{\sqrt{3}}}
F\left(g\left(\e^{-\frac{\varphi}{\sqrt{3}}}\right)\right) = \frac{A}{F_R(A)} - \frac{F(A)}{F_R(A)^2}\, .
\end{align}
Here $A$ corresponds to the scalar curvature in the original Jordan frame and $g\left(\e^{-\frac{\varphi}{\sqrt{3}}}\right)$ is obtained by solving
the equation $\varphi = -\frac{1}{\sqrt{3}}\ln F_R(A)$ with respect to $A=g\left(\e^{-\frac{\varphi}{\sqrt{3}}}\right)$.

For the scalar-tensor theory with the potential $V(\varphi)$ for the scalar field $\varphi$, which may be identified with the inflaton field,
the slow-roll parameters $\epsilon$ and $\eta$ are defined by follows,
\begin{align}
\label{cr26}
\epsilon \equiv \frac{1}{2} \left( \frac{V'(\varphi)}{V(\varphi)} \right)^2 \, , \quad
\eta \equiv \frac{V''(\varphi)}{V(\varphi)} \, .
\end{align}
The above expressions could be used even in the anisotropic universe.
By using the slow-roll indexes $\epsilon$ and $\eta$, we can express
the observational indices $n_s$ and $r$ as follows,
\begin{align}
\label{cr27}
n_s -1 = - 6\epsilon + 2 \eta\, , \quad r=16\epsilon\, .
\end{align}
In the case that we can neglect the contributions from the matter, the slow-roll parameters $\epsilon$ and $\eta$ and
the observational indices $n_s$ and $r$ do not change in the frame \cite{Nojiri:2022ski}.
Therefore we work in the Einstein frame.

Because
\begin{align}
\label{Avarphi}
\frac{d}{d\varphi}=\frac{dA}{d\varphi}\frac{d}{dA} = -\frac{\sqrt{3} F_R(A)}{F_{RR}(A)} \frac{d}{dA}\, ,
\end{align}
we find
\begin{align}
\label{Vdashes}
V'(\varphi)= \sqrt{3} \left( \frac{A}{F_R(A)} - \frac{2 F(A)}{F_R(A)^2} \right) \, , \quad
V''(\varphi)= 3 \left( \frac{1}{F_{RR}(A)} + \frac{A}{F_R(A)} - \frac{4F(A)}{F_R(A)^2} \right) \, ,
\end{align}
and therefore, by using (\ref{cr26}), we obtain \cite{Bamba:2014wda},
\begin{align}
\label{cr26B}
\epsilon =
\frac{3}{2} \left( \frac{AF_R(A) - 2 F(A)}{AF_R(A) - F(A)} \right)^2
\, , \quad
\eta =
\frac{3 \left( F_R(A)^2 + A F_{RR}(A) F_R(A) - 4 F_{RR}(A) F(A) \right)}{F_{RR}(A) \left( A F_R(A) - F(A) \right)} \, .
\end{align}
We also obtain the expressions of the observational indices $n_s$ and $r$ in (\ref{cr27}),
\begin{align}
\label{cr27B}
n_s -1 =&\, - 9 \left( \frac{AF_R(A) - 2 F(A)}{AF_R(A) - F(A)} \right)^2
+ \frac{6 \left( F_R(A)^2 + A F_{RR}(A) F_R(A) - 4 F_{RR}(A) F(A) \right)}{F_{RR}(A) \left( A F_R(A) - F(A) \right)} \, , \nonumber \\
r=&\, 24 \left( \frac{AF_R(A) - 2 F(A)}{AF_R(A) - F(A)} \right)^2\, .
\end{align}

As an example and just for simplicity by choosing $F_2=0$, we consider the model (\ref{FRB17}) with (\ref{FRB20}),
\begin{align}
\label{FRB17BB}
F(R) = - \frac{54}{7} {H_0}^2 F_1 + \frac{9}{7} F_1 R - \frac{F_1}{84 {H_0}^2} R^2 \, ,
\end{align}
where $R$ behaves as $R \sim 12{H_0}^2 + \frac{C^2 \e^{-6H_0 t}}{{F_1}^2}$ in (\ref{FRB15}).
Then in the late time $t\to \infty$, we find
\begin{align}
\label{FRB17BB}
F(R) \sim 6 {H_0}^2 F_1 + \frac{C^2 \e^{-6H_0 t}}{F_1}\, ,\quad
F_R(R) \sim F_1  - \frac{C^2 \e^{-6H_0 t}}{42 {H_0}^2 F_1} \, , \quad
F_{RR}(R) = - \frac{F_1}{42 {H_0}^2} \, ,
\end{align}
and therefore
\begin{align}
\label{cr27B2}
n_s -1 \sim 9 \, , \quad
r\sim \frac{54 C^4 \e^{-12H_0 t}}{49{F_1}^4} \, .
\end{align}
The obtained value of $n_s - 1$ might not be realistic and we may need to consider more realistic model.

\section{$F(R)$ Gravity With an Extra Scalar Field and Anisotropic Inflation}

Now let us consider anisotropic inflation in the case of $F(R)$
gravity with an extra scalar field, in which case we consider the
following gravitational action,
\begin{align}
\label{CS5} S_\phi = \frac{1}{2\kappa^2} \int d^4 x \sqrt{-g}
\left[ F(R) - \frac{\omega(\phi)}{2}\partial_\mu \phi \partial^\mu
\phi - V(\phi) \right] \, ,
\end{align}
the variation of which with respect to the scalar field $\phi$,
yields,
\begin{align}
\label{FCS5S}
0 = \nabla^\mu \left( \omega(\phi) \partial_\mu \phi \right)
 - \frac{\omega'(\phi)}{2}\partial_\mu \phi \partial^\mu \phi - V'(\phi) \, .
\end{align}
Assuming that $\phi=t$, Eqs.~(\ref{FRB1}), (\ref{FRB3}), and
(\ref{FRB4}) can be rewritten as follows,
\begin{align}
\label{FRBp1}
&\, \frac{1}{2}\omega + V
 - \frac{1}{2} F +\left( 3 \dot H + 3 H^2  + \sum_{i=1}^3 \left({\dot\beta}^i\right)^2 \right) F_R - 3 H \frac{d F_R}{dt} = - \kappa^2 \rho \, , \\
\label{FRBp3}
&\, \frac{1}{2}\omega - V + \frac{1}{2} F  -\left(  \dot H + 3 H^2 \right) F_R + \frac{d^2 F_R}{dt^2} + 2H \frac{d F_R}{dt} = - \kappa^2 p \, , \\
\label{FRBp4}
&\, -\left( {\ddot\beta}^i + 3 H {\dot\beta}^i\right) F_R - {\dot\beta}^i \frac{d F_R}{dt} = 0 \, ,
\end{align}
We should note that Eq.~(\ref{FRBp4}) is identical to
Eq.~(\ref{FRB4}), which could indicate that the isometry is mainly
controlled by the functional form of $F(R)$. Eqs.~(\ref{FRBp1})
and (\ref{FRBp3}) can be combined and can be rewritten as follows,
\begin{align}
\label{FRBp5}
\omega =&\, - \left( 2 \dot H + \sum_{i=1}^3 \left({\dot\beta}^i\right)^2 \right) F_R + H \frac{d F_R}{dt} -  \frac{d^2 F_R}{dt^2} - \kappa^2 \left(\rho + p\right)\, , \\
\label{FRBp6}
V =&\, \frac{1}{2} F  -\left( 2 \dot H + 3 H^2 - \frac{1}{2}\sum_{i=1}^3 \left({\dot\beta}^i\right)^2 \right) F_R + \frac{5}{2}H \frac{d F_R}{dt}
+ \frac{1}{2} \frac{d^2 F_R}{dt^2} + \frac{\kappa^2}{2} \left( - \rho + p \right) \, .
\end{align}
As in Eq. (\ref{FRB8}), we may assume that the matter is composed
by perfect fluids with constant EoS parameters $w_i$, which yields
the scale factor $a$-dependence of $\rho$ and $p$, $\rho=\rho(a)$
and $p=p(a)$. Then if we consider an arbitrary expansion of the
Universe, with an arbitrary scale factor $a=a(t)$ the
time-dependence of the right hand sides of Eqs. (\ref{FRBp5}) and
(\ref{FRBp6}), can be obtained and therefore we find the
time-dependence of the kinetic term function $\omega$ and of the
scalar potential $V$, namely $\omega(t)$ and $V(t)$. By replacing
$t$ in $\omega(t)$ and $V(t)$ with the scalar field $phi$, we find
the $\phi$-dependence of $\omega$ and $V$,
$\omega(\phi)=\omega(t=\phi)$ and $V(\phi) = V(t=\phi)$ and we can
specify the model realizing the given scale factor $a=a(t)$.

As a special case we consider the Einstein gravity where $F(R)=R$.
In the case of the Einstein gravity, Eqs.~(\ref{FRBp4}), (\ref{FRBp5}), and (\ref{FRBp6})  have the following forms,
\begin{align}
\label{FRBE4}
&\,  {\ddot\beta}^i + 3 H {\dot\beta}^i = 0 \, , \\
\label{FRBp5}
\omega =&\,  2 \dot H + \sum_{i=1}^3 \left({\dot\beta}^i\right)^2 - \kappa^2 \left(\rho + p\right)\, , \\
\label{FRBp6}
V =&\,  \dot H + 3 H^2 + \frac{1}{2}\sum_{i=1}^3 \left({\dot\beta}^i\right)^2 + \frac{\kappa^2}{2} \left( - \rho + p \right) \, .
\end{align}
Eq.~(\ref{FRBE4}) can be integrated to give,
\begin{align}
\label{Ea} \beta^i = {\tilde C}^i a^{-3}\, ,
\end{align}
where ${\tilde C}^i$'s are constants of the integration. As in Eq.
(\ref{FRB6}), Eq.~(\ref{an2}) indicates that,
\begin{align}
\label{FRB6E}
0 = \sum_{i=1}^3 {\tilde C}^i \, .
\end{align}
From Eq.~(\ref{Ea}) we can see that the anisotropy disappears in
the late Universe where the scale factor $a$ is large, even in the
scalar-Einstein theory. By using Eq. (\ref{Ea}),
Eqs.~(\ref{FRBp5}) and (\ref{FRBp6}) can be rewritten as follows,
\begin{align}
\label{FRBp5B}
\omega =&\,  2 \dot H + {\tilde C}^2 a^{-6} - \kappa^2 \left(\rho + p\right)\, , \\
\label{FRBp6B} V =&\,  \dot H + 3 H^2 + {\tilde C}^2 a^{-6}  +
\frac{\kappa^2}{2} \left( - \rho + p \right) \, ,
\end{align}
where in this case as in Eq. (\ref{FRB8}), we define ${\tilde
C}^2$ as follows,
\begin{align}
\label{FRB8E}
{\tilde C}^2 \equiv \sum_{i=1} \left({\tilde C}^i\right)^2 \, .
\end{align}

Let us now consider an example, having to do with the inflationary
epoch. During inflation, the contribution from the perfect matter
fluids could be neglected, and therefore we put $\rho=p=0$. We may
assume,
\begin{align}
\label{ex1} a(t) = \frac{1}{\e^{-H_0\left( t - t_0 \right)} +
\left( \frac{t_0}{t} \right)^n}\, ,
\end{align}
where $H_0$, $t_0$, and $n$ are constants and we assume that $H_0$
and $n$ are positive. When $t\ll t_0$, $a(t)$ behaves as an
exponential function $a\sim \e^{H_0 t}$, which corresponds to the
de Sitter expansion and therefore the era $t\ll t_0$ can be
regarded to be the inflationary era. On the other hand, when $t\gg
t_0$, the scale factor $a(t)$ behaves as a power function of $t$,
$a(t)\sim \left( \frac{t}{t_0} \right)^n$. Therefore, the era
$t\gg t_0$ can be identified with the post-inflationary era. For
the scale factor $a$ in (\ref{ex1}), by using Eq. (\ref{FRB6E}),
we find the behavior of $\beta^i(t)$ as follows,
\begin{align}
\label{ex2}
\beta^i = {\tilde C}^i \left( \e^{-H_0\left( t - t_0 \right)} + \left( \frac{t_0}{t} \right)^n \right)^3\, ,
\end{align}
which becomes small in the late Universe and therefore the universe becomes isometric since the anisotropies get smoothed
away. For the scale factor $a$ in Eq. (\ref{ex1}) the Hubble rate
is,
\begin{align}
\label{ex3}
H =&\, \frac{H_0 \e^{-H_0\left( t - t_0 \right)} + \frac{n}{t_0} \left( \frac{t_0}{t} \right)^{n+1} }{\e^{-H_0\left( t - t_0 \right)} + \left( \frac{t_0}{t} \right)^n} \, , \nonumber \\
\dot H =&\, \frac{- {H_0}^2 \e^{-H_0\left( t - t_0 \right)} - \frac{n(n+1)}{{t_0}^2} \left( \frac{t_0}{t} \right)^{n+2}
+ \left( H_0 \e^{-H_0\left( t - t_0 \right)} + \frac{n}{t_0} \left( \frac{t_0}{t} \right)^{n+1} \right)^2 }
{\left( \e^{-H_0\left( t - t_0 \right)} + \left( \frac{t_0}{t} \right)^n\right)^2} \nonumber \\
=&\, \frac{ 2 H_0 \e^{-H_0\left( t - t_0 \right)} \frac{n}{t_0} \left( \frac{t_0}{t} \right)^{n+1} - \frac{n(n+1)}{{t_0}^2} \left( \frac{t_0}{t} \right)^{n+2}
+ \frac{n^2}{{t_0}^2} \left( \frac{t_0}{t} \right)^{2n+2}} {\left( \e^{-H_0\left( t - t_0 \right)} + \left( \frac{t_0}{t} \right)^n\right)^2} \, ,
\end{align}
thus Eq.~(\ref{FRBp5B}) gives,
\begin{align}
\label{FRBp5B}
\omega (\phi) =&\, \frac{ \frac{4 n H_0}{t_0} \e^{-H_0\left( \phi - t_0 \right)} \left( \frac{t_0}{\phi} \right)^{n+1} -  \frac{2n(n+1)}{{t_0}^2} \left( \frac{t_0}{\phi} \right)^{n+2}
+ \frac{2n^2}{{t_0}^2} \left( \frac{t_0}{\phi} \right)^{2n+2}} {\left( \e^{-H_0\left( \phi - t_0 \right)} + \left( \frac{t_0}{\phi} \right)^n\right)^2}
+ {\tilde C}^2  \left( \e^{-H_0\left( \phi - t_0 \right)} + \left( \frac{t_0}{\phi} \right)^n \right)^6 \, , \\
\label{FRBp6B}
V (\phi) =&\,  \frac{ 3{H_0}^2 \e^{-2 H_0\left( \phi - t_0 \right)}
+ 8 H_0 \e^{-H_0\left( \phi - t_0 \right)} \frac{n}{t_0} \left( \frac{t_0}{\phi} \right)^{n+1} - \frac{n(n+1)}{{t_0}^2} \left( \frac{t_0}{\phi} \right)^{n+2}
+ \frac{4 n^2}{{t_0}^2} \left( \frac{t_0}{\phi} \right)^{2n+2}} {\left( \e^{-H_0\left( \phi - t_0 \right)} + \left( \frac{t_0}{\phi} \right)^n\right)^2} \, ,
\end{align}
which specify the model.

In the late time $t=\phi\to +\infty$, the expressions in
(\ref{FRBp5B}) and (\ref{FRBp6B}) are reduced as
\begin{align}
\label{FRBp5C}
\omega (\phi) \sim -  \frac{2n(n+1)}{{t_0}^2} \left( \frac{t_0}{\phi} \right)^{-n+2} \, , \quad
V (\phi) \sim - \frac{n(n+1)}{{t_0}^2} \left( \frac{t_0}{\phi} \right)^{-n+2}
\end{align}
Because $\omega (\phi)$ is negative if $\phi$ is positive, the
scalar field becomes a ghost, and therefore the model could be
regarded as a kind of effective theory.\footnote{ In order to
avoid this problem, we may consider only the case that $n$ is an
odd integer and $\phi$ is negative, and furthermore, we identify
$\phi=-t$ instead of $\phi=t$.} We now normalize the scalar field
by
\begin{align}
\label{norm}
\varphi \equiv \int d\phi \sqrt{ - \omega (\phi)} \sim 2 \sqrt{\frac{n+1}{n}}\left( \frac{\phi}{t_0} \right)^\frac{n}{2}
= 2 \sqrt{\frac{n+1}{n}}\left( \frac{t}{t_0} \right)^\frac{n}{2} \, ,
\end{align}
and we find $V(\varphi) \propto \varphi^{2 - \frac{4}{n}}$.
Then if we define the slow-roll parameters by (\ref{cr26}) and the observational indices by (\ref{cr27}), we find
\begin{align}
\label{cr26CC}
\epsilon \sim&\, \frac{1}{2\varphi^2} \left( 2 - \frac{4}{n} \right)^2 \, , \quad
\eta \sim \frac{1}{\varphi^2} \left( 2 - \frac{4}{n} \right)\left( 1 - \frac{4}{n} \right) \, , \nonumber \\
n_s -1 \sim&\, - \frac{1}{\varphi^2}  \left( 2 - \frac{4}{n} \right)\left( 4 - \frac{4}{n} \right) \, , \quad
r \sim \frac{8}{\varphi^2} \left( 2 - \frac{4}{n} \right)^2 \, .
\end{align}
By using (\ref{norm}), we find
\begin{align}
\label{cr26CC2}
\epsilon \sim&\, \left( 2 - \frac{4}{n} \right)^2 \frac{n}{8\left( n + 1 \right)} \left( \frac{t}{t_0} \right)^{-n} \, , \quad
\eta \sim \left( 2 - \frac{4}{n} \right)\left( 1 - \frac{4}{n} \right) \frac{n}{4\left( n + 1 \right)} \left( \frac{t}{t_0} \right)^{-n} \, , \nonumber \\
n_s -1 \sim&\, - \left( 2 - \frac{4}{n} \right)\left( 4 - \frac{4}{n} \right) \frac{n}{4\left( n + 1 \right)} \left( \frac{t}{t_0} \right)^{-n} \, , \quad
r \sim \left( 2 - \frac{4}{n} \right)^2 \frac{2n}{\left( n + 1 \right)} \left( \frac{t}{t_0} \right)^{-n} \, .
\end{align}

\section{Anisotropic Inflation in Modified Gauss-Bonnet Gravity}

In this section, we shall study anisotropic inflation in the
context of another mainstream modified gravity, namely for
modified Gauss-Bonnet gravity (MGB), for Bianchi IX cosmology. We
start with the action of the MGB-gravity and cast it into a
convenient formulation using auxiliary scalar degrees of freedom,
and then specialize it to the Bianchi IX cosmology. We will also
consider two additional cosmological models, related with the
Bianchi IX geometry: the Bianchi I and the Taub Cosmology. The aim
of this section is to derive the Lagrangian and the equation of
motion for these models.

Let us consider a class of modified gravity theories where an
arbitrary function of the topological Gauss-Bonnet term is added
to the Lagrangian of GR.

The action of the modified Gauss-Bonnet gravity \cite{reviews1} is
the following:
\begin{align}
    \label{AnG1}
    S=\int d^4x\sqrt{-g}\left( \frac{1}{2k^2}R +f(G) \right) \, ,
\end{align}
where $G$ is the Gauss-Bonnet invariant, defined as:
\begin{align}
    G=R^2-4R_{\mu \nu}R^{\mu \nu}+R_{\mu \nu \alpha \beta}R^{\mu \nu \alpha \beta} \, .
    \label{AnG2}
\end{align}
The action \eqref{AnG1} can be rewritten in a convenient form by
introducing two auxiliary fields $A$ and $B$, as follows,
\begin{align}
    S=\int d^4x\sqrt{-g}\left( \frac{1}{2k^2}R + B(G-A)+f(A)\right) \, .
    \label{AnG3}
\end{align}
Upon varying the action with respect to the auxiliary scalar
degrees of freedom $B$ we obtain $A=G$, and by substituting this
into Eq. \eqref{AnG3}, the action of Eq. \eqref{AnG1} is recovered.
On the other hand, by varying the action with respect to $A$ we
get $B=f'(A)$. Hence,
\begin{align}
    S=\int d^4x\sqrt{-g}\left( \frac{1}{2k^2}R f'(A)G-Af'(A)+f(A)\right) \,.
    \label{AnG4}
\end{align}
Renaming the variable $A=\phi$ and $f'(A)=\xi(\phi)$,
$V(\phi)=Af'(A)+f(A)$, and so the action \eqref{AnG4} becomes,
\begin{align}
    S=\int d^4x\sqrt{-g}\left( \frac{1}{2k^2}R - \xi(\phi)G - V(\phi) \right) \, .
    \label{AnG5}
\end{align}
We want to describe the features of the Universe adopting the
Bianchi IX cosmology, which is the most general geometry allowed
by the homogeneity constraint. The relevance of the dynamics of
Bianchi models consists in the role these geometries could have
played in a very primordial Universe, before the inflationary
phase. For a quantum discussion on the Bianchi IX model see
\cite{Wilson-Ewing:2017vju, Antonini:2018gdd}. The model has been
extensively studied in GR by C.Misner
\cite{Misner:1969hg,Misner:1969ae} but few analysis were
made in modified Gauss-Bonnet gravity.

The line element of a generic Bianchi model \cite{Landau:1975pou}
is,
\begin{align}
ds^{2}=-N^{2}(t)dt^{2}+a^{2}(t)(\omega^{l})^{2}+b^{2}(t)(\omega^{m})^{2}+c^{2}(t)(\omega^{n})^{2} \, .
\label{AnG6}
\end{align}
where the functions $\omega^{i}$ are the 1-forms that define the
geometry of the model. For the Bianchi I and Bianchi IX
cosmologies the 1-forms are:
\begin{align}
\begin{array}{ll}
\omega^{l}=dx,     &  \omega^{l}=\cos\psi d\theta+\sin\psi\sin\theta d\phi,\\
\omega^{m}=dy,     &  \omega^{m}=\sin\psi d\theta-\cos\psi\sin\theta d\phi, \\
\omega^{n}=dz,     &  \omega^{n}=d\psi+\cos\theta d\phi. \\
\end{array}
\label{AnG7}
\end{align}
We can calculate the Gauss-Bonnet invariant, the Ricci Scalar and
the $R_{\mu \nu}R^{\mu \nu}$ for the Bianchi IX cosmology using
the following relationships,
\begin{align}
    R_0^0=-\frac{1}{2}\frac{\partial}{\partial t}\chi_{\alpha}^{\alpha}-\frac{1}{4}\chi_{\alpha}^{\beta}\chi_{\beta}^{\alpha},  \quad
    R_{\alpha}^{\beta}=-P_{\alpha}^{\beta}-\frac{1}{2\sqrt{\gamma}}\frac{\partial}{\partial t}\left( \sqrt{\gamma}\chi_{\alpha}^{\beta} \right).
    \label{AnG8}
\end{align}
where $P_{\alpha}^{\beta}$ is the 3-dimensional Ricci scalar and
$\gamma$ is the determinant of the spatial metric. The Riemann
tensor will be computed using the relationship for the Riemann
tensor in the tetrad formalism,
\begin{align}
    R_{a b c d}=\gamma_{abc,d}-\gamma_{abd,c}+\gamma_{abf}\left( \gamma^f_{cd}-\gamma^f_{dc}\right)+\gamma_{afc}\gamma^f_{bd}-\gamma_{afd}\gamma^f_{bc} \, ,
    \label{AnG9}
\end{align}
where $g_{ik}=\eta_{ab}e_i^{(a)}e_k^{(b)}$ and the
$\gamma_{acb}=e_{(a)i};ke^i_{(b)}e^k_{(c)}$ are the Ricci rotation
coefficient. We can introduce their linear combinations
$\lambda_{abc}=\gamma_{abc}-\gamma_{acb}$. For the Bianchi IX
cosmology we can express them in terms of the structure constants
and the time derivative of the spatial metric as follows,
\begin{align}
    \begin{array}{cc}
    \lambda^a_{bc}=C^a_{bc}\, ,     & \lambda^0_{ab}=\chi_{ab}\, .\\
    \end{array}
    \label{AnG10}
\end{align}
The non-vanishing components of the Riemann tensor are,
\begin{align}
    \begin{array}{ccc}
    R_{0101} & R_{0202} & R_{0303}\\
    R_{1212} & R_{1313} & R_{2323}\\
    R_{0123} & R_{0231} & R_{0312}\\
    \end{array}
    \label{AnG11}
\end{align}
and their permutations allowed by the Bianchi identities. We can
now compute the Gauss-Bonnet invariant, which has the following
form,
\begin{align}
        G =&8 \left\{ \right. \frac{\Ddot{a}}{a}\left[ \frac{\Dot{b}}{b}\frac{\Dot{c}}{c} - \frac{1}{2}U_1\right] + \frac{\Ddot{b}}{b}\left[ \frac{\Dot{a}}{a}\frac{\Dot{c}}{c} - \frac{1}{2}U_2\right] + \frac{\Ddot{c}}{c}\left[ \frac{\Dot{a}}{a}\frac{\Dot{b}}{b} - \frac{1}{2}U_3\right] +
         \frac{1}{2}\left(\frac{\Dot{a}}{a}\right)^2\frac{\partial U_1}{\partial \log{a}} + \frac{1}{2}\left(\frac{\Dot{b}}{b}\right)^2\frac{\partial U_2}{\partial \log{b}} + \frac{1}{2}\left(\frac{\Dot{c}}{c}\right)^2\frac{\partial U_3}{\partial \log{c}} + \nonumber\\
        & - \frac{1}{2}\frac{\Dot{a}}{a}\frac{\Dot{b}}{b}\left[\frac{\partial U_1}{\partial \log{a}}+\frac{\partial U_2}{\partial \log{b}}-\frac{\partial U_3}{\partial \log{c}} \right] - \frac{1}{2}\frac{\Dot{a}}{a}\frac{\Dot{c}}{c}\left[\frac{\partial U_1}{\partial \log{a}}-\frac{\partial U_2}{\partial \log{b}}+\frac{\partial U_3}{\partial \log{c}} \right] - \frac{1}{2}\frac{\Dot{b}}{b}\frac{\Dot{c}}{c}\left[-\frac{\partial U_1}{\partial \log{a}}+\frac{\partial U_2}{\partial \log{b}}+\frac{\partial U_3}{\partial \log{c}} \right]\left.\right\},
\label{AnG12}
\end{align}
where $U_1,U_2$ and $U_3$ are functions defined for the Bianchi I
model,
\begin{equation}
    U_1=U_2=U_3=0 \, ,
    \label{AnG13}
\end{equation}
and for the Bianchi IX model:
\begin{align}
        &U_1 = \frac{-3a^4+b^4+c^4+2a^2b^2+2a^2c^2-2b^2c^2}{2a^2b^2c^2} \,, \nonumber \\
        &U_2 = \frac{a^4-3b^4+c^4+2a^2b^2-2a^2c^2+2b^2c^2}{2a^2b^2c^2} \,,\nonumber \\
        &U_3 = \frac{a^4+b^4-3c^4-2a^2b^2+2a^2c^2+2b^2c^2}{2a^2b^2c^2} \,.\nonumber \\
\label{AnG14}
\end{align}
The Lagrangian of the Bianchi IX model assumes the following form,
\begin{align}
    \mathcal{L}_{IX} &=  -a b c V(\phi) +2\Dot{a}\Dot{b}c + 2\Dot{a}b\Dot{c} + 2a\Dot{b}\Dot{c} +\frac{a^4 + b^4 + c^4 - 2 a^2 b^2  - 2 a^2 c^2 - 2 b^2 c^2 }{2 a b c} + \nonumber \\
    & + 8\xi '(\phi ) \Dot{\phi}\left( \right.\Dot{a} \Dot{b} \Dot{c}-\frac{3 a^2 \Dot{a}}{4 b c}+\frac{c^3 \Dot{a}}{4 a^2 b}+\frac{c \Dot{a}}{2 b}+\frac{b \Dot{a}}{2 c}+\frac{b^3 \Dot{a}}{4 a^2 c} -\frac{b c \Dot{a}}{2 a^2}+\frac{a^3 \Dot{b}}{4 b^2 c}+\frac{a \Dot{b}}{2 c}-\frac{a c \Dot{b}}{2 b^2} + \nonumber\\
    &+\frac{c^3 \Dot{b}}{4 a b^2}+\frac{c \Dot{b}}{2 a}-\frac{3 b^2 \Dot{b}}{4 a c} +\frac{a^3 \Dot{c}}{4 b c^2}+\frac{a \Dot{c}}{2 b}-\frac{a b \Dot{c}}{2 c^2}+\frac{b \Dot{c}}{2 a}+\frac{b^3 \Dot{c}}{4 a c^2}-\frac{3 c^2 \Dot{c}}{4 a b} \left. \right)\, .
\label{AnG15}
\end{align}
The equations of motion can be found using the field equations for
the theory at hand, and we eventually get,
\begin{align}
        &0 = -\frac{b^3}{2 a^2 c}+\frac{b c}{a^2}-\frac{c^3}{2 a^2 b}+\frac{3 a^2}{2 b c}-2 c \Ddot{b}-2 \Dot{b} \Dot{c}-2 b \Ddot{c}-\frac{b}{c}  -\frac{c}{b} -bcV(\phi)+ \Dot{\phi} +\xi '(\phi) \left(\frac{6 b^2 \Dot{b}}{a^2 c}-\frac{6 b^2 \Dot{b}}{a^2 c}-8 \Ddot{b} \Dot{c}-8 \Dot{b} \Ddot{c}\right) + \nonumber \\
        & +\Dot{\phi}^2 \xi ''(\phi ) \left(-\frac{2 b^3}{a^2 c}+\frac{4 b c}{a^2}-\frac{2 c^3}{a^2 b}+\frac{6 a^2}{b c}-8 \Dot{b} \Dot{c}-\frac{4 b}{c}-\frac{4 c}{b}\right) + \Ddot{\phi} \xi '(\phi) +\left(-\frac{2 b^3}{a^2 c}+\frac{4 b c}{a^2}-\frac{2 c^3}{a^2 b}+\frac{6 a^2}{b c}-8 \Dot{b}\Dot{c}-\frac{4 b}{c}-\frac{4 c}{b}\right); \nonumber\\
        &\nonumber \\
        &0= a b c V'(\phi ) + \xi '(\phi)G\, .
    \label{AnG16}
\end{align}
The equations of motion for $b(t)$ and $c(t)$ can be found cycling
the scale factors and $G$ is defined in Eq. \eqref{AnG12}. For the
sake of simplicity we considered the Taub cosmological model
\cite{Taub:1950ez}, with the following line element,
\begin{align}
ds^{2}=-N^{2}(t)dt^{2}+a^{2}(t)\left[(\omega^{l})^{2}+b(t)(\omega^{m})^{2}\right]+c^{2}(t)(\omega^{n})^{2}\,,
\label{AnG17}
\end{align}
so, the action \eqref{AnG15} becomes,
\begin{align}
        \mathcal{L}_{Taub}=& \frac{b^3}{4 a^2}+2 a \Dot{a} \Dot{b}+b \Dot{a}^2-b -a^2 b V(\phi) +\Dot{\phi}\xi'(\phi) \left( -\frac{6 b^2 \Dot{b}}{a^2}+8 \Dot{a}^2\Dot{b} + \frac{4 b^3 \Dot{a}}{a^3}+8 \Dot{b} \right) \, ,
    \label{AnG18}
\end{align}
and the equations of motion are:,
\begin{align}
       & 2 b \Ddot{a}+2 \Dot{a} \Dot{b}+2 a\Ddot{b} +2 a b V(\phi)+\frac{b^3}{2 a^3} - \xi''(\phi)\left( -16 \Dot{a} \Dot{b} \Dot{\phi}^2-\frac{4 b^3 \Dot{\phi}^2}{a^3} \right)-\xi'(\phi)\left( -16 \Ddot{a} \Dot{b} \Dot{\phi}-16 \Dot{a} \Ddot{b} \Dot{\phi}-16 \Dot{a}\Dot{b} \Ddot{\phi}-\frac{4 b^3 \Dot{\phi}}{a^3} \right)=0;  \nonumber\\
       &\nonumber\\
       & a^2 V(\phi) - \frac{3 b^2}{4 a^2}+1 +2 a \Ddot{a}+\Dot{a}^2  -\xi''(\phi)\left( -8 \Dot{a}^2 \Dot{\phi}^2+\frac{6 b^2 \Dot{\phi}^2}{a^2}-8 \Dot{\phi}^2 \right) -\xi'(\phi)\left(-8 \Dot{a}^2 \Ddot{\phi}-16 \Dot{a}\Ddot{a}\Dot{\phi}+\frac{6 b^2 \Ddot{\phi}}{a^2}-8\Ddot{\phi} +  \right)=0;\nonumber \\
       &\nonumber\\
       &\xi'(\phi)\left( \frac{2 b^3 \Ddot{a}}{a^3}+4 \Dot{a}^2 \Ddot{b}+\frac{12 b^2 \Dot{a} \Dot{b}}{a^3}-\frac{6 b^3 \Dot{a}^2}{a^4}+8 \Dot{a} \Ddot{a} \Dot{b}-\frac{3 b^2 \Ddot{b}}{a^2}+\right.\left.-\frac{6 b \Dot{b}^2}{a^2}+4 \Ddot{b} \right)+ a^2 b V'(\phi)= 0.
       \label{AnG19}
\end{align}
where the first two are dynamical equations for the metric $a(t)$
and $b(t)$ and the last one is the equation for the scalar field
$\phi(t)$.

\subsection{Theory Reconstruction in Taub Cosmology and Specific
Cosmological Solutions}

We can use the equation of motion \eqref{AnG19} to find the
potentials $\xi(\phi)$ and $V(\phi)$ which lead to interesting
cosmological solutions. The reconstruction of the corresponding
theory is a well known procedure in the flat FRW geometry
\cite{Nojiri:2010wj, Nojiri:2008nt} and here we try to extend
these results to the Taub cosmological model. For the sake of
simplicity, we will derive the potentials $V(\phi)$ and
$\xi'(\phi)=\partial \xi(\phi)/\partial \phi$, because only the
derivative of this potential appears in the action \eqref{AnG18}.

\subsubsection{Anisotropic de Sitter Solution}

An interesting solution of the Einstein field equations is the de
Sitter one, which describes an empty Universe with a cosmological
constant. The line element of this solution is:
\begin{equation}
    ds^2 = dt^2 - a_0^2\cosh{\left(H_0 t \right)} d\Omega_3 \, ,
    \label{Tr4}
\end{equation}
where we are dealing with a non-stationary metric in a synchronous
reference frame. We can identify $a(t) = a_0\cosh{\left(H_0 t
\right)}$ and bring back this line element to the FRW one with $K
= 1$. This form of the metric represents a de Sitter solution in
an isotropic closed Universe. The de Sitter model describes a
singularity free Universe, which collapses to a finite minimum of
the volume and then re-expands.

In this subsection we assume de Sitter solutions for the expansion
factors $a(t),b(t)$ and find the family of potentials
$V(\phi),\xi'(\phi)$ which lead to these cosmological solutions.
The scalar field has a power-law dependence on time and the
cosmological model is still anisotropic, thus,
\begin{align}
    \begin{array}{cc}
    a(t)=a_0\cosh{H_a t}, & b(t)=b_0\cosh{H_b t}\, ,  \\
    \end{array}
    \label{Tr5}
\end{align}
and the scalar field is,
\begin{align}
    \phi(t)=t^k \, .
    \label{Tr6}
\end{align}
We can integrate the equations of motion \eqref{AnG19} and obtain
the following potentials,
\begin{align}
        V\left(\phi\right) = & -\frac{\text{sech}^4(\phi_a)}{8 \text{a}_0^4} \left[3 \text{a}_0^4 \text{H}_a^2 \cosh (4 \phi_a)+ \right.
        4 \text{a}_0^2 +4 \left(2 \text{a}_0^4 \text{H}_a^2+\text{a}_0^2\right) \cosh (2 \phi_a) + \left. -3 \text{b}_0^2\left( \cosh (2 \phi_b) +1\right)-3 \text{b}_0^25 \text{a}_0^4 \text{H}_a^2 \right]+\nonumber\\
        &+ c_1\exp{\left(f_1\left(\phi\right) - \sqrt{\frac{f_2\left(\phi\right)}{f_3\left(\phi\right)} + \frac{f_4\left(\phi\right)}{f_3\left(\phi\right)}} \right)^2}  +c_2\exp{\left(f_1\left(\phi\right) + \sqrt{\frac{f_2\left(\phi\right)}{f_3\left(\phi\right)} + \frac{f_4\left(\phi\right)}{f_3\left(\phi\right)}}\right)^2}, \nonumber\\
        &\nonumber \\
        \xi'\left(\phi\right) = & \frac{\text{a}_0^4 \cosh ^4(\phi_a) \cosh (\phi_a)}{f_5\left(\phi\right)}V\left(\phi\right).
        \label{Tr7}
\end{align}
with $\phi_a = H_a\phi^{1/k}$. The explicit form of the auxiliary
functions $f_1,f_2,f_3,f_4$ and $f_5$ is given in Appendix A.

\subsubsection{Isotropic de Sitter Solution}

Now let us seek for potentials which lead to an isotropic de
Sitter solution for a closed Universe,
\begin{align}
    \begin{array}{c}
    a(t)=b(t)=r_0\cosh{(H_0t)}\, ,
    \end{array}
    \label{Tr8}
\end{align}
Let us consider a simple power-law dependence of the scalar field
on time:
\begin{equation}
    \phi(t)=t^k \,,
    \label{Tr9}
\end{equation}
then we can integrate the equations of motion to find the
potentials,
\begin{align}
        V(\phi)=& -3 {H_0}^2 e^{-\tilde{\phi}} \tan ^{-1}\left(e^{\tilde{\phi}}\right) +3 {H_0}^2 e^{\tilde{\phi}} \tan ^{-1}\left(e^{\tilde{\phi}}\right) -\frac{3}{4 r_0^2} + \frac{3 e^{-\tilde{\phi}} \tan ^{-1}\left(e^{\tilde{\phi}}\right)}{4 r_0^2}+\frac{e^{3 \tilde{\phi}}}{2 H_0 \left(e^{2 \tilde{\phi}}+1\right)}+
        \nonumber\\
        &+e^{-2 \tilde{\phi}} \text{sech}\left(\tilde{\phi}\right)-\frac{3 e^{\tilde{\phi}} \tan ^{-1}\left(e^{\tilde{\phi}}\right)}{4 r_0^2},\nonumber\\
        &\nonumber\\
         \xi'(\phi) = & \frac{ \tan ^{-1}\left(e^{\tilde{\phi}}\right) \cosh ^3\left(\tilde{\phi}\right)}{4 H_0 f\left(\phi \right)}  -\frac{H_0 r_0^2  \sinh \left(\tilde{\phi}\right) \cosh \left(\tilde{\phi}\right)}{2 f\left(\phi\right)}   - \frac{H_0 r_0^2  \tan ^{-1}\left(e^{\tilde{\phi}}\right) \cosh ^3\left(\tilde{\phi}\right)}{f\left(\phi\right)} + \frac{ \sinh \left(\tilde{\phi}\right) \cosh \left(\tilde{\phi}\right)}{8 H_0 f\left( \phi\right)} + \nonumber\\
        & +\frac{r_0^2  e^{2 \tilde{\phi}} \sinh \left(\tilde{\phi}\right)}{24 {H_0}^2 f\left(\phi\right)} + \frac{r_0^2  e^{-2 \tilde{\phi}} \cosh \left(\tilde{\phi}\right)}{3 H_0 f\left(\phi\right)}    -\frac{r_0^2  e^{2 \tilde{\phi}} \cosh \left(\tilde{\phi}\right)}{12 {H_0}^2 f\left(\phi\right)} +\frac{r_0^2  e^{-2 \tilde{\phi}} \sinh \left(\tilde{\phi}\right)}{6 H_0 f\left(\phi\right)},\nonumber\\
    \label{Tr10}
\end{align}
with,
\begin{equation}
    f\left(\phi\right) = \frac{4 {H_0}^2 k r_0^2 \sinh ^2\left(\tilde{\phi}\right)+k}{\left(\phi ^{1/k}\right)^{1-k}}, \quad  \tilde{\phi} = H_0 \phi ^{1/k}.
\end{equation}
We can perturb the solutions \eqref{Tr8} and \eqref{Tr9} adding a
small perturbation to the expansion factor and to the scalar
field, as follows,
\begin{align}
        a(t)=r_0\cosh(H_0t)+\delta a_1(t)\, , \quad
        \phi(t)=t^k+\delta \phi_1(t)\, ,
    \label{Tr11}
\end{align}
with $\delta$ is a small parameter. We can study the evolution of
the perturbation using the potentials \eqref{Tr10} and the
equations of motion \eqref{AnG19}. Since $a(t)=b(t)$, the first two
equations in \eqref{AnG19} take the same form and the equations of
motions have the following form,
\begin{align}
        &0=-4 \delta  H_0 r_0 a_1'(t) \sinh \left(\tilde{t}\right)-4 \delta  {H_0}^2 r_0 a_1(t) \cosh \left(\tilde{t}\right) -4 \delta  r_0 a_1(t) V(t) \cosh \left(\tilde{t}\right)-2 r_0^2 V(t) \cosh ^2\left(\tilde{t}\right)+\\
        &-3 {H_0}^2 r_0^2 \cosh \left(2 \tilde{t}\right)-{H_0}^2 r_0^2-\frac{1}{2}-4 \delta  r_0 a_1''(t) \cosh \left(\tilde{t}\right) + -16 H_0^3 k r_0^2 t^{k-1} \sinh \left(2 \tilde{t}\right)-4 k^2 t^{k-2}+4 k t^{k-2} \nonumber\\
        &+\xi ''(t) \left\{ 4 t^{-1}\left[ 8 \delta  H_0 k r_0 t^k a_1'(t) \sinh \left(\tilde{t}\right) \right. \right.\left. \left. +4 {H_0}^2 r_0^2 \sinh ^2\left(\tilde{t}\right) \left(4 \delta ^2 t \phi_1'(t)^2+k t^k\right)+k t^k \right]  \right\} + \nonumber\\
        &32 \xi '(t) \delta H_0 \left\{ -  k r_0 t^{k-1} a_1''(t) \sinh \left(\tilde{t}\right) \right. -  k^2 r_0 t^{k-2} a_1'(t) \sinh \left(\tilde{t}\right)+ k r_0 t^{k-2} a_1'(t) \sinh \left(\tilde{t}\right) \left. \right\}+ \nonumber\\
        &-32 \delta  {H_0}^2 k r_0 t^{k-1} a_1'(t) \cosh \left(\tilde{t}\right)-16 \delta  {H_0}^2 r_0^2 \phi_1''(t) \sinh ^2\left(\tilde{t}\right)-4 \delta  \phi_1''(t)-16 \delta  H_0^3 r_0^2 \phi_1'(t) \sinh \left(2 \tilde{t}\right) + \nonumber\\
        &-16 {H_0}^2 k^2 r_0^2 t^{k-2} \sinh ^2\left(\tilde{t}\right) +16 {H_0}^2 k r_0^2 t^{k-2} \sinh ^2\left(\tilde{t}\right)+ \nonumber\\
        &\nonumber\\
        &0=-6 \xi '(t) \left[ {H_0}^2 r_0  4 \delta  r_0 a_1''(t) \sinh ^2\left(\tilde{t}\right)+\cosh \left(\tilde{t}\right)\right]+\delta  a_1''(t)+4 \delta  H_0^3 r_0^2 a_1'(t) \sinh \left(2 \tilde{t}\right) +2 {H_0}^4 r_0^3 \sinh \left( \tilde{t}\right) \sinh \left( 2 \tilde{t} \right) \nonumber\\
        &-r_0^3 k^{-1} t^{1-k} V'(t) \cosh ^3\left( \tilde{t} \right) -\delta  r_0^2 t^{1-2 k} V'(t) \cosh ^2\left( \tilde{t} \right) \left( 3 k a_1(t) t^k -r_0 t \cosh \left( \tilde{t} \right) \phi _1'(t) \right)k^{-2},
    \label{Tr12}
\end{align}
where we considered only the terms linear in $\delta$ and
$\tilde{t}= H_0 t $. We can solve the second equation in
\eqref{Tr12} with respect to $\phi_1'(t)$ and substitute it into
the first equation to obtain a third order differential equation
for $a_1(t)$. The resulting equation is,
\begin{align}
    F_3 a_1'''(t) + F_2 a_1''(t) + F_1 a_1'(t) + F_0 a_1(t) =0
    \label{Tr13}
\end{align}
with,
\begin{align}
        F_0 =& \frac{2 \delta  r_0 e^{- \tilde{t}} \left(2 H_0^3 \left(e^{2 \tilde{t}}+1\right){}^3+4 H_0 e^{3 \tilde{t}}+e^{3 \tilde{t}}\right)}{H_0 \left(e^{2 \tilde{t}}+1\right){}^2}, \nonumber\\
        & \nonumber\\
        F_1 =& \delta  H_0 r_0 e^{-2 \tilde{t}} \left(e^{\tilde{t}} \left(e^{2 \tilde{t}}-1\right)+3 \left(e^{2 \tilde{t}}+1\right){}^2 \tan ^{-1}\left(e^{\tilde{t}}\right)\right) -\frac{3 \delta  e^{-2 \tilde{t}} \left(e^{\tilde{t}} \left(e^{2 \tilde{t}}-1\right)+\left(e^{2 \tilde{t}}+1\right){}^2 \tan ^{-1}\left(e^{\tilde{t}}\right)\right)}{4 H_0 r_0}+ \nonumber\\
        &+\frac{\delta  r_0 e^{2 \tilde{t}} \left(e^{2 \tilde{t}}+3\right)}{2 {H_0}^2 \left(e^{2 \tilde{t}}+1\right)} - \frac{2 \delta  r_0 \left(-e^{-2 \tilde{t}}-3\right)}{H_0 \left(e^{2 \tilde{t}}+1\right)}, \nonumber\\
        &\nonumber\\
        F_2 =& \frac{\delta  e^{-2 \tilde{t}} \left(e^{2 \tilde{t}}+1\right) \left(e^{\tilde{t}}+\left(e^{2 \tilde{t}}-1\right) \tan ^{-1}\left(e^{\tilde{t}}\right)\right)}{2 {H_0}^2 r_0}  -2 \delta  r_0 e^{-2 \tilde{t}} \left(e^{2 \tilde{t}}+1\right) \left(2 e^{\tilde{t}}+\left(e^{2 \tilde{t}}-1\right) \tan ^{-1}\left(e^{\tilde{t}}\right)\right) + \nonumber\\
        &-\frac{\delta  r_0 e^{2 \tilde{t}} \left(2 e^{2 \tilde{t}}+e^{4 \tilde{t}}+3\right)}{3 H_0^3 \left(e^{2 \tilde{t}}+1\right){}^2}  -\frac{4 \delta  r_0 e^{-2 \tilde{t}} \left(2 e^{2 \tilde{t}}+3 e^{4 \tilde{t}}+1\right)}{3 {H_0}^2 \left(e^{2 \tilde{t}}+1\right){}^2},\nonumber \\
        &\nonumber\\
        F_3  =& \frac{\delta  e^{-2 \tilde{t}} \left(e^{\tilde{t}} \left(e^{2 \tilde{t}}-1\right)+\left(e^{2 \tilde{t}}+1\right){}^2 \tan ^{-1}\left(e^{\tilde{t}}\right)\right)}{4 H_0^3 r_0}  -\frac{\delta  r_0 e^{2 \tilde{t}} \left(e^{2 \tilde{t}}+3\right)}{6 {H_0}^4 \left(e^{2 \tilde{t}}+1\right)} + \frac{2 \delta  r_0 e^{-2 \tilde{t}} \left(3 e^{2 \tilde{t}}+1\right)}{3 H_0^3 \left(e^{2 \tilde{t}}+1\right)} + \nonumber\\
        & \frac{\delta  r_0 e^{-2 \tilde{t}} \left(e^{\tilde{t}} \left(e^{2 \tilde{t}}-1\right)+\left(e^{2 \tilde{t}}+1\right){}^2 \tan ^{-1}\left(e^{\tilde{t}}\right)\right)}{H_0}.
    \label{Tr14}
\end{align}
This equation can be solved numerically with different parameters
($H_0, r_0$) and initial conditions $a_1(0), a_1'(0)$ and
$a_1''(0)$.

\subsection{Scalar Potentials in Bianchi I Cosmology}

The Gauss-Bonnet invariant for the Bianchi I cosmology makes the
potentials $U_1,U_2$ and $U_3$ in Eq.\eqref{AnG12} vanish, obtaining
the following expression,
\begin{align}
    G_I=8\left( \Ddot{a}\Dot{b}\Dot{c}+\Dot{a}\Ddot{b}\Dot{c}+\Dot{a}\Dot{b}\Ddot{c}\right)
    \label{BI1}
\end{align}
and so, the Lagrangian of the cosmological model becomes,
\begin{align}
    \mathcal{L}_{I}= & 2 c \Dot{a} \Dot{b}+2 b \Dot{a} \Dot{c}+2 a \Dot{b} \Dot{c} + 8\Dot{\phi} \xi'(\phi) \Dot{a} \Dot{b} \Dot{c} -a b c V(\phi).
    \label{BI2}
\end{align}
Using the field equations we found the following equation of
motion,
\begin{align}
        &-8 \Dot{b} \Dot{c} \Dot{\phi}^2 \xi''(\phi )+\Dot{b} \Dot{c}+\xi'(\phi ) \left(-8 \Ddot{b} \Dot{c} \Dot{\phi}-8 \Dot{b} \Ddot{c} \Dot{\phi}-8 \Dot{b} \Dot{c} \Ddot{\phi}\right)+\frac{1}{2} \left(-2 c \Ddot{b}-4 \Dot{b} \Dot{c}-2 b \Ddot{c}\right)-b c V(\phi )=0,\nonumber\\
        & -8 \Dot{a} \Dot{c} \Dot{\phi}^2 \xi''(\phi )+\Dot{a} \Dot{c}+\xi'(\phi ) \left(-8 \Ddot{a} \Dot{c} \Dot{\phi}-8 \Dot{a} \Ddot{c} \Dot{\phi}-8 \Dot{a} \Dot{c} \Ddot{\phi}\right)+\frac{1}{2} \left(-2 c \Ddot{a}-4 \Dot{a} \Dot{c}-2 a \Ddot{c}\right)-a c V(\phi )=0, \nonumber\\
        &-8 \Dot{a} \Dot{b} \Dot{\phi}^2 \xi''(\phi )+\Dot{a} \Dot{b}+\xi'(\phi ) \left(-8 \Ddot{a} \Dot{b} \Dot{\phi}-8 \Dot{a} \Ddot{b} \Dot{\phi}-8 \Dot{a} \Dot{b} \Ddot{\phi}\right)+\frac{1}{2} \left(-2 b \Ddot{a}-4 \Dot{a} \Dot{b}-2 a \Ddot{b}\right)-a b V(\phi )=0, \nonumber \\
        &\xi '(\phi ) \left(-8 \Ddot{a} \Dot{b} \Dot{c}-8 \Dot{a} \Ddot{b} \Dot{c}-8 \Dot{a} \Dot{b} \Ddot{c}\right)-a b c V'(\phi)=0.
        \label{BI3}
\end{align}
The simplest solution to the Einstein equation in the framework of
the Bianchi classification is the Kasner one, a solution for an
empty, anisotropic and flat Universe. In this case the spatial
line element reduces to,
\begin{align}
    dl^2 = t^{2p_1}\left( dx^1 \right) + t^{2p_2}\left( dx^2 \right) + t^{2p_3}\left( dx^3 \right),
    \label{BI4}
\end{align}
where $p_1, p_2$ and $p_3$ are the Kasner indices and they satisfy
two relations,
\begin{align}
    \begin{array}{cc}
     p_1 + p_2 + p_3 =1 \\
     p_1^2 + p_2^2 + p_3^2 = 1
    \end{array}
    \label{BI5}
\end{align}
and so there is only one independent parameter that describes the
solution. We are looking for the form of potential which leads to
Kasner solutions,
\begin{align}
    \begin{array}{ccc}
    a(t)=t^{p_1}, & b(t)=t^{p_2}, & c(t)=t^{p_3}. \\
    \end{array}
    \label{BI6}
\end{align}
We assume a simple power-law dependence of the scalar field on
time,
\begin{align}
    \phi(t)=t^k,
    \label{BI7}
\end{align}
and so we can solve the equations of motion \label{BI3} to find
the following potentials,
\begin{align}
    V(\phi) =& -\frac{\text{p}_1 \left(\text{p}_2^2+\text{p}_2 (\text{p}_3-1)+(\text{p}_3-1) \text{p}_3\right) \phi ^{-2/\text{k}} E_1}{\text{p}_1^2+\text{p}_1 (E_1-\text{p}_1)+2 (\text{p}_2+\text{p}_3-1)} + c_1 \left(\phi ^{E_3/2\text{k}}\right) + c_2 \left(\phi ^{\overline{E_3}/2\text{k}}\right)\, ,\nonumber\\
    &\nonumber\\
    \xi'(\phi)=& \frac{\left(\phi ^{1/\text{k}}\right)^{2-\text{k}}}{16 \text{p}_2 \text{p}_3 \text{k}}\left(  -\frac{4 \left(\text{p}_2^2+\text{p}_2 (\text{p}_3-1)+(\text{p}_3-1) \text{p}_3\right)}{\text{p}_1^2+\text{p}_1 (E_1-\text{p}_1)+2 (\text{p}_2+\text{p}_3-1)}
        +  \frac{i \sqrt{\text{p}_1 E_1 E_2}  +\text{p}_2+\text{p}_3+1}{\text{p}_1 E_1} \right) + \nonumber \\
    &+c_1 \left(\phi ^{1/2\text{k}}\right)^{\left(-i \sqrt{\text{p}_1 E_1 E_2}  -\text{p}_2-\text{p}_3+3\right)} + c_2 \left(\phi ^{1/2\text{k}}\right)^{ \left(+i \sqrt{\text{p}_1 E_1 E_2} -\text{p}_2-\text{p}_3+3\right)}\, ,
    \label{BI8}
\end{align}
with $E_1=\text{p}_1+\text{p}_2+\text{p}_3-3$, $E_2 = -\frac{4 \text{p}_1^2+4 \text{p}_1 (\text{p}_2+\text{p}_3-3)+(\text{p}_2+\text{p}_3+1)^2}{\text{p}_1 E_1}$
and $E_3 = i\sqrt{\text{p}_1 E_1 E_2} -\text{p}_2-\text{p}_3-1$.

\section{Conclusions and Future Perspective}

In this work we considered some anisotropic evolution scenarios in
the context of modified gravity in several of its various form.
These scenarios can be relevant for pre-inflationary eras and may
leave their imprint on the four dimensional classical effective
inflationary theory. We considered several mainstream modified
gravities, such as $F(R)$ gravity with an extra scalar field and
in its Jordan frame vacuum form, and also Gauss-Bonnet gravities.
In all the cases we considered, the field equations can be
utilized as a reconstruction method and several evolutions of
specific form like Taub and Bianchi Universes can be realized.
Using these reconstruction techniques, we were able to present the
formalism of studying anisotropic inflation in several modified
gravities. This work fills in the gap in the literature between
ordinary isotropic inflationary theories in modified gravity with
a single scale factor and anisotropic inflation in modified
gravity theories. Our aim was to present reconstruction techniques
to be utilized in order to obtain phenomenological outcomes out of
these theories. This is not an easy task, however with this work
we made the first step in the literature towards this direction.
We also answered the vital question whether cosmological solutions
of elevated importance exist, like for example isotropic or
anisotropic de Sitter solutions. Finally, for the Gauss-Bonnet
theories, we provided expressions for the scalar potentials, in
the case of a Bianchi I cosmology. Admittedly, the way towards
obtaining phenomenological information for these modified gravity
theories in the context of anisotropic inflation, is long and
thorny, however we presented the first step in this direction, the
formalization of anisotropic inflation in non-trivial modified
gravity theories. Now what remains as a future perspective of this
work is to quantitatively check the actual effects of the
anisotropies on the effective inflationary theory starting at the
first horizon crossing when the primordial modes exit the horizon
for the first time. The short wavelength modes will directly be
probed by future gravitational wave experiments like the LISA and
DECIGO, and thus we will have a direct grasp on these  modes.
Therefore, if the Universe was anisotropic pre-inflationary, these
evolution conditions may set and determine the initial conditions
of the Universe at the pre-inflationary stages. Of course as the
Universe enters the inflationary era, the anisotropies will be
smoothed away, however, the anisotropies may affect directly the
initial conditions of the inflationary epoch, for example the
scale of inflation. These issues should be separately considered
in a future work, based on the formalism we discussed in this
paper.

\section*{Acknowledgments}

This work was supported by MINECO (Spain), project
PID2019-104397GB-I00 (S.D.O). This work by S.D.O was also
partially supported by the program Unidad de Excelencia Maria de
Maeztu CEX2020-001058-M, Spain.


\begin{thebibliography}{99}






\bibitem{inflation1}
 A.~D.~Linde,
 Lect.\ Notes Phys.\ {\bf 738} (2008) 1
 [arXiv:0705.0164 [hep-th]].

\bibitem{inflation2} D.~S.~Gorbunov and V.~A.~Rubakov,
``Introduction to the theory of the early universe: Cosmological
perturbations and inflationary theory,'' Hackensack, USA: World
Scientific (2011) 489 p;
%


\bibitem{inflation3}A.~Linde,
arXiv:1402.0526 [hep-th];


\bibitem{inflation4}D.~H.~Lyth and A.~Riotto,
Phys.\ Rept.\  {\bf 314} (1999) 1 [hep-ph/9807278].


\bibitem{Planck:2018vyg}
N.~Aghanim \textit{et al.} [Planck],
Astron. Astrophys. \textbf{641} (2020), A6 [erratum: Astron.
Astrophys. \textbf{652} (2021), C4]
doi:10.1051/0004-6361/201833910 [arXiv:1807.06209 [astro-ph.CO]].


\bibitem{reviews1}
 S.~Nojiri, S.~D.~Odintsov and V.~K.~Oikonomou,
  Phys.\ Rept.\  {\bf 692} (2017) 1
  [arXiv:1705.11098 [gr-qc]].

\bibitem{reviews2}


 S. Capozziello, M. De Laurentis,
   Phys.\ Rept.\  {\bf 509}, 167 (2011);\\
 V.~Faraoni and S.~Capozziello,
  Fundam.\ Theor.\ Phys.\  {\bf 170} (2010).



\bibitem{reviews3}
S. Nojiri, S.D. Odintsov,
  eConf {\bf C0602061}, 06 (2006)
  [Int.\ J.\ Geom.\ Meth.\ Mod.\ Phys.\  {\bf 4}, 115 (2007)].


   \bibitem{reviews4}

S. Nojiri, S.D. Odintsov,
   Phys.\ Rept.\  {\bf 505}, 59 (2011);




\bibitem{reviews5}

G.~J.~Olmo,
  Int.\ J.\ Mod.\ Phys.\ D {\bf 20} (2011) 413
  [arXiv:1101.3864 [gr-qc]].



\bibitem{Nojiri:2003ft}
S.~Nojiri and S.~D.~Odintsov,
Phys.\ Rev.\ D {\bf 68} (2003) 123512
doi:10.1103/PhysRevD.68.123512 [hep-th/0307288].



\bibitem{Appleby:2009uf}
S.~A.~Appleby, R.~A.~Battye and A.~A.~Starobinsky,
JCAP \textbf{06} (2010), 005 doi:10.1088/1475-7516/2010/06/005
[arXiv:0909.1737 [astro-ph.CO]].


\bibitem{Artymowski:2014gea}
M.~Artymowski and Z.~Lalak,
JCAP \textbf{09} (2014), 036 doi:10.1088/1475-7516/2014/09/036
[arXiv:1405.7818 [hep-th]].



\bibitem{Yashiki:2020naf}
M.~Yashiki, N.~Sakai and R.~Saito,
Phys. Rev. D \textbf{102} (2020) no.4, 043504
doi:10.1103/PhysRevD.102.043504 [arXiv:2004.00864 [astro-ph.CO]].


\bibitem{Ellis:2020krl}
J.~Ellis, D.~V.~Nanopoulos, K.~A.~Olive and S.~Verner,
JCAP \textbf{03} (2021), 052 doi:10.1088/1475-7516/2021/03/052
[arXiv:2008.09099 [hep-ph]].


\bibitem{Chen:2022zkc}
H.~Chen, T.~Katsuragawa and S.~Matsuzaki,
[arXiv:2206.02130 [gr-qc]].









\bibitem{Nojiri:2007as}
S.~Nojiri and S.~D.~Odintsov,
Phys.\ Lett.\ B {\bf 657} (2007) 238
doi:10.1016/j.physletb.2007.10.027 [arXiv:0707.1941 [hep-th]].

\bibitem{Nojiri:2007cq}
S.~Nojiri and S.~D.~Odintsov,
Phys.\ Rev.\ D {\bf 77} (2008) 026007
doi:10.1103/PhysRevD.77.026007 [arXiv:0710.1738 [hep-th]].

\bibitem{Cognola:2007zu}
G.~Cognola, E.~Elizalde, S.~Nojiri, S.~D.~Odintsov, L.~Sebastiani
and S.~Zerbini,
Phys.\ Rev.\ D {\bf 77} (2008) 046009
doi:10.1103/PhysRevD.77.046009 [arXiv:0712.4017 [hep-th]].

\bibitem{Nojiri:2006gh}
S.~Nojiri and S.~D.~Odintsov,
Phys.\ Rev.\ D {\bf 74} (2006) 086005
doi:10.1103/PhysRevD.74.086005 [hep-th/0608008].

\bibitem{Appleby:2007vb}
S.~A.~Appleby and R.~A.~Battye,
Phys.\ Lett.\ B {\bf 654} (2007) 7
doi:10.1016/j.physletb.2007.08.037 [arXiv:0705.3199 [astro-ph]].



\bibitem{Elizalde:2010ts}
E.~Elizalde, S.~Nojiri, S.~D.~Odintsov, L.~Sebastiani and
S.~Zerbini,
Phys.\ Rev.\ D {\bf 83} (2011) 086006
doi:10.1103/PhysRevD.83.086006 [arXiv:1012.2280 [hep-th]].


\bibitem{Odintsov:2020nwm}
S.~D.~Odintsov and V.~K.~Oikonomou,
Phys. Rev. D \textbf{101} (2020) no.4, 044009
doi:10.1103/PhysRevD.101.044009 [arXiv:2001.06830 [gr-qc]].


\bibitem{Sa:2020fvn}
P.~M.~S\'a,
Phys. Rev. D \textbf{102} (2020) no.10, 103519
doi:10.1103/PhysRevD.102.103519 [arXiv:2007.07109 [gr-qc]].



\bibitem{Oikonomou:2022wuk}
V.~K.~Oikonomou and I.~Giannakoudi,
[arXiv:2205.08599 [gr-qc]].


\bibitem{Odintsov:2021urx}
S.~D.~Odintsov and V.~K.~Oikonomou,
Phys. Lett. B \textbf{824} (2022), 136817
doi:10.1016/j.physletb.2021.136817 [arXiv:2112.02584 [gr-qc]].



\bibitem{Bamonti:2021jmg}
N.~Bamonti, A.~Costantini and G.~Montani,
[arXiv:2103.17063 [gr-qc]].






\bibitem{Chen:2021nkf}
C.~B.~Chen and J.~Soda,
JCAP \textbf{09} (2021), 026 doi:10.1088/1475-7516/2021/09/026
[arXiv:2106.04813 [hep-th]].



\bibitem{Do:2021lyf}
T.~Q.~Do and W.~F.~Kao,
Eur. Phys. J. C \textbf{81} (2021) no.6, 525
doi:10.1140/epjc/s10052-021-09334-y [arXiv:2104.14100 [gr-qc]].



\bibitem{Sadeghi:2021egp}
J.~Sadeghi and S.~Noori Gashti,
Eur. Phys. J. C \textbf{81} (2021) no.4, 301
doi:10.1140/epjc/s10052-021-09103-x [arXiv:2104.00117 [hep-th]].



\bibitem{Hiramatsu:2020jes}
T.~Hiramatsu, K.~Murai, I.~Obata and S.~Yokoyama,
JCAP \textbf{03} (2021), 047 doi:10.1088/1475-7516/2021/03/047
[arXiv:2008.03233 [astro-ph.CO]].



\bibitem{Do:2020ler}
T.~Q.~Do, W.~F.~Kao and I.~C.~Lin,
Eur. Phys. J. C \textbf{81} (2021) no.5, 390
doi:10.1140/epjc/s10052-021-09181-x [arXiv:2003.04266 [gr-qc]].





\bibitem{Gong:2019hwj}
J.~O.~Gong, T.~Noumi, G.~Shiu, J.~Soda, K.~Takahashi and
M.~Yamaguchi,
JCAP \textbf{08} (2020), 027 doi:10.1088/1475-7516/2020/08/027
[arXiv:1910.11533 [hep-th]].





\bibitem{Fujita:2018zbr}
T.~Fujita, I.~Obata, T.~Tanaka and S.~Yokoyama,
JCAP \textbf{07} (2018), 023 doi:10.1088/1475-7516/2018/07/023
[arXiv:1801.02778 [astro-ph.CO]].





\bibitem{Ito:2017bnn}
A.~Ito and J.~Soda,
Eur. Phys. J. C \textbf{78} (2018) no.1, 55
doi:10.1140/epjc/s10052-018-5534-5 [arXiv:1710.09701 [hep-th]].




\bibitem{Lahiri:2016jqv}
S.~Lahiri,
JCAP \textbf{09} (2016), 025 doi:10.1088/1475-7516/2016/09/025
[arXiv:1605.09247 [hep-th]].





\bibitem{Ito:2015sxj}
A.~Ito and J.~Soda,
Phys. Rev. D \textbf{92} (2015) no.12, 123533
doi:10.1103/PhysRevD.92.123533 [arXiv:1506.02450 [hep-th]].





\bibitem{Blanco-Pillado:2015dfa}
J.~J.~Blanco-Pillado and M.~Minamitsuji,
JCAP \textbf{06} (2015), 024 doi:10.1088/1475-7516/2015/06/024
[arXiv:1501.07427 [hep-th]].


\bibitem{Naruko:2014bxa}
A.~Naruko, E.~Komatsu and M.~Yamaguchi,
JCAP \textbf{04} (2015), 045 doi:10.1088/1475-7516/2015/04/045
[arXiv:1411.5489 [astro-ph.CO]].




\bibitem{Emami:2013bk}
R.~Emami and H.~Firouzjahi,
JCAP \textbf{10} (2013), 041 doi:10.1088/1475-7516/2013/10/041
[arXiv:1301.1219 [hep-th]].




\bibitem{Watanabe:2010bu}
M.~a.~Watanabe, S.~Kanno and J.~Soda,
Mon. Not. Roy. Astron. Soc. \textbf{412} (2011), L83-L87
doi:10.1111/j.1745-3933.2011.01010.x [arXiv:1011.3604
[astro-ph.CO]].



\bibitem{Kanno:2010nr}
S.~Kanno, J.~Soda and M.~a.~Watanabe,
JCAP \textbf{12} (2010), 024 doi:10.1088/1475-7516/2010/12/024
[arXiv:1010.5307 [hep-th]].



\bibitem{Watanabe:2010fh}
M.~a.~Watanabe, S.~Kanno and J.~Soda,
Prog. Theor. Phys. \textbf{123} (2010), 1041-1068
doi:10.1143/PTP.123.1041 [arXiv:1003.0056 [astro-ph.CO]].



\bibitem{Dulaney:2010sq}
T.~R.~Dulaney and M.~I.~Gresham,
Phys. Rev. D \textbf{81} (2010), 103532
doi:10.1103/PhysRevD.81.103532 [arXiv:1001.2301 [astro-ph.CO]].



\bibitem{Barrow:2009gx}
J.~D.~Barrow and S.~Hervik,
Phys. Rev. D \textbf{81} (2010), 023513
doi:10.1103/PhysRevD.81.023513 [arXiv:0911.3805 [gr-qc]].



\bibitem{Rothman:1986gg}
T.~Rothman and G.~F.~R.~Ellis,
Phys. Lett. B \textbf{180} (1986), 19-24
doi:10.1016/0370-2693(86)90126-7


\bibitem{Belinsky:1982pk}
V.~a.~Belinsky, I.~m.~Khalatnikov and E.~m.~Lifshitz,
Adv. Phys. \textbf{31} (1982), 639-667
doi:10.1080/00018738200101428

\bibitem{Wilson-Ewing:2017vju}
E.~Wilson-Ewing,
Class. Quant. Grav. \textbf{35} (2018) no.6, 065005
doi:10.1088/1361-6382/aaab8b
[arXiv:1711.10943 [gr-qc]].


\bibitem{Antonini:2018gdd}
S.~Antonini and G.~Montani,
Phys. Lett. B \textbf{790} (2019), 475-483
doi:10.1016/j.physletb.2019.01.050
[arXiv:1808.01304 [gr-qc]].


\bibitem{Misner:1969hg}
C.~W.~Misner,
Phys. Rev. Lett. \textbf{22} (1969), 1071-1074
doi:10.1103/PhysRevLett.22.1071


\bibitem{Misner:1969ae}
C.~W.~Misner,
Phys. Rev. \textbf{186} (1969), 1319-1327
doi:10.1103/PhysRev.186.1319


\bibitem{Landau:1975pou}
L.~D.~Landau and E.~M.~Lifschits,

\bibitem{Taub:1950ez}
A.~H.~Taub,
Annals Math. \textbf{53} (1951), 472-490
doi:10.2307/1969567


\bibitem{Nojiri:2010wj}
S.~Nojiri and S.~D.~Odintsov,
Phys. Rept. \textbf{505} (2011), 59-144
doi:10.1016/j.physrep.2011.04.001
[arXiv:1011.0544 [gr-qc]].


\bibitem{Nojiri:2008nt}
S.~Nojiri and S.~D.~Odintsov,
TSPU Bulletin \textbf{N8(110)} (2011), 7-19
[arXiv:0807.0685 [hep-th]].




\bibitem{Nojiri:2022ski}
S.~Nojiri, S.~D.~Odintsov and V.~K.~Oikonomou,
Nucl. Phys. B \textbf{980} (2022), 115850
doi:10.1016/j.nuclphysb.2022.115850
[arXiv:2205.11681 [gr-qc]].

\bibitem{Bamba:2014wda}
K.~Bamba, S.~Nojiri, S.~D.~Odintsov and D.~S\'aez-G\'omez,
Phys. Rev. D \textbf{90} (2014), 124061
doi:10.1103/PhysRevD.90.124061
[arXiv:1410.3993 [hep-th]].



\end{thebibliography}
\end{document}